\documentclass[aps,prd,twocolumn,superscriptaddress,eqsecnum,nofootinbib,showpacs,altaffilletter]{revtex4-1}

\usepackage{graphicx}
\usepackage{euscript,amssymb}
\usepackage{amsfonts}
\usepackage{amsmath}
\usepackage{amssymb}
\usepackage{graphicx}
\usepackage{euscript}
\usepackage{color}

\newcommand{\ben}{\begin{eqnarray}}
\newcommand{\een}{\end{eqnarray}}
\newcommand{\bn}{\begin{equation}\label}
\newcommand{\be}{\begin{equation}}
\newcommand{\ee}{\end{equation}}
\newcommand{\ba}{\begin{eqnarray}}
\newcommand{\ea}{\end{eqnarray}}
\newcommand{\n}{\label}

\newcommand{\ga}{\gamma}
\newcommand{\ro}{\rho}
\newcommand{\ep}{\epsilon}

\begin{document}

\title{$k$-essence in the DGP brane-world cosmology}

\author{Mariam Bouhmadi-L\'{o}pez}
\email{mariam.bouhmadi@ist.utl.pt}
\affiliation{Centro Multidisciplinar de Astrof\'{\i}sica - CENTRA, Departamento de F\'{\i}sica, Instituto Superior T\'ecnico, Av. Rovisco Pais 1,
1049-001 Lisboa, Portugal}
\author{Luis Chimento}\email{chimento@df.uba.ar}
\affiliation{Departamento de Física, Facultad de Ciencias Exactas y Naturales, Universidad de Buenos Aires, Ciudad Universitaria, Pabellón I, 1428 Buenos Aires, Argentina}

\bibliographystyle{plain}


\begin{abstract}
We analyse a DGP brane filled with a $k$-essence field and assume the $k$-field evolving linearly with the cosmic time of the brane. We then solve analytically the Friedmann equation and deduce the different  behaviour of the brane at the low and the high energy regimes.  The asymptotic behaviour can be quite different involving accelerating branes, big bangs, big crunches, big rips or quiescent singularities. The latter correspond to a type of sudden singularity.
\end{abstract}

\date{\today}
\maketitle

\section{Introduction}

One of the most puzzling discovery of the last years in physics is the current acceleration of the universe \cite{Perlmutter:1998np}. Despite the huge efforts made so far to find a well motivated theoretical framework for this behaviour no consensus has been reached, even though a fine tuned cosmological constant is the simplest option to match the current observations \cite{Perlmutter:1998np,Spergel:2003cb,Cole:2005sx,Tegmark:2003uf}. There are two main streams of thought that try to explain the late-time speed up of the universe:  (i) modified theories of gravity on large scale, which by weakening the gravitational interaction on those scales allow inflationary universes (cf. Refs.~\cite{Nojiri:2006ri,Capozziello:2007ec,Sotiriou:2008rp,Durrer:2008in,DeFelice:2010aj}) (ii) a dark energy component, corresponding to a new component on the cosmic pie of the universe, that violates the strong energy condition and therefore allows  accelerating universes \cite{Copeland:2006wr}.

One approach to build a modified theory of gravity relies on extra-dimensions. The idea of extra-dimensions is quite old in physics and it dates back to Kaluza and Klein. On the last decades it has been invoked by string theory as an approach to unify the different interactions in nature and in particular as a new road to obtain a consistent theory of quantum gravity. On this modern approach not only the extra-dimensions plays a crucial role but also the branes \cite{Maartens:2010ar}. In particular, within the context of brane-world models, our universe corresponds to a brane; i.e. a 4-dimensional (4d) hypersuface, embedded in a higher dimensional space-time dubbed the bulk. The simplest of these models are Randall-Sundrum models corresponding to an ultra-violet modification of general relativity (GR) \cite{Randall:1999ee} or  the Dvali-Gabadadze-Porrati (DGP) model corresponding to an infra-red modification of GR \cite{dgp}. The latter is a promising approach, despite its shortcomings \cite{Koyama:2007za}, to describe the current inflationary epoch of the universe.

The DGP model contains two set of solutions, usually referred to as the self-accelerating branch and the normal branch. While the self-accelerating branch as it name indicates is accelerating in the absence of any kind of dark energy, the normal branch requires some sort of stuff to describe any inflationary era of the universe. On the other hand, the normal branch is free from the ghost issue present on the self-accelerating DGP solution. However, it is important to highlight that there is a duality between both branches. Indeed, we can obtain the homogeneous and isotropic solutions and investigate them in anyone of the two branches (see section~\ref{sec2}). In the present paper, we show how it is possible to get a set of accelerating branes of the normal branch by means of a $k$-essence field embedded on the brane \cite{Armendariz-Picon:1999rj,Chimento:2009hg}. A complementary motivation for our analysis is to classify the different kind of singularities, in particular those related to dark energy, that may show up on the brane.

The paper is outlined as follows. In section II, we present the model we will analyse, i.e. a homogeneous and isotropic DGP brane filled with a $k$-essence field that evolves linearly with the cosmic time of the brane. At the end of this section we introduce the form invariant transformation, that preserve the form of the Friedmann and conservation equations, to show that there is a duality between the solutions of the normal branch and the self-accelerating one in the DGP model. This duality relates expanding and contracting solutions among themselves. In section III, we present the analytical solutions of the model for the normal branch and analyse the different behaviour of the brane depending on the equation of state of the $k$-field and its energy density.  On the last section, we summarise our main results and conclude.

\section{The DGP model with a $k$-essence field}
\label{sec2}

We will analyse the evolution of a DGP brane \cite{dgp} filled with  a $k$-essence field \cite{Armendariz-Picon:1999rj} confined on the brane.  The energy momentum tensor reads \cite{Armendariz-Picon:1999rj}
\begin{equation}
T_{\mu\nu}=V(\phi)(2F_x\phi_\mu\phi_\nu-g_{\mu\nu}F), \quad F_x=\frac{dF}{dx}.
\label{Tmunuk}
\end{equation}
Here $F=F(x)$ is an arbitrary function of the kinetic energy of the $k$-field $x=g^{\mu\nu}\nabla_\mu\phi\nabla_\nu\phi$ and $V(\phi)$ is a potential.
We will assume a spatially flat, homogeneous and isotropic brane, therefore, the modified Friedmann equation reads \cite{dgpfriedmann}
\begin{equation}
H^2+\frac{\epsilon}{r_c} H=\frac{\kappa_4^2}{3}\rho,
\label{Friedmann}
\end{equation}
where $r_c$ is the cross-over scale and is related to the ratio between the 4-dimensional effective gravitational constant, $\kappa_4^2$, and the 5-dimensional gravitational constant of the bulk. Finally, $\epsilon=\pm 1$ stands for the two branches of the DGP model, $\epsilon=-1$ corresponds to the self-accelerating DGP branch and $\epsilon=+1$ corresponds to the normal DGP branch \cite{dgpfriedmann}. 
For latter convenience we rewrite the Friedmann equation as
\begin{equation}
H=\frac{\epsilon}{2r_c}\left[-1\pm\sqrt{1+\frac{\rho}{\rho_m}}\right],
\label{Friedmannroot}
\end{equation}
where 
\begin{equation}
\rho_m=\frac{3}{4 r_c^2 \kappa_4^2}.
\label{rhom}
\end{equation}
The energy momentum tensor is conserved on the brane and therefore
\begin{equation}
\dot{\rho}+3H(\rho+p)=0.
\label{conservation}
\end{equation}
In addition, it can be shown that the Raychaudurui equation can be written as
\begin{equation}
\dot H= \frac12 \kappa_4^2(p+\rho)\left(\frac{\epsilon-\sqrt{1+\frac{\rho}{\rho_m}}}{\sqrt{1+\frac{\rho}{\rho_m}}}\right).
\label{Raychaudhuri}
\end{equation}

From Eq.~(\ref{Tmunuk}), we obtain the energy density $\ro$ and pressure $p$ of the $k$-field
\begin{equation} 
\rho=V(\phi)(F-2xF_x), \qquad p=-V(\phi)F,
\label{defrhop}
\end{equation} 
with equation of state $\ga=(p+\ro)/\ro=-2xF_x/(F-2xF_x)$. The evolution equation of the $k$-field can be deduced by substituting Eqs.~(\ref{defrhop}) into Eq.~(\ref{conservation})
\begin{equation}
(F_x+2xF_{xx})\ddot\phi+3HF_x\dot\phi+\frac{V'}{2V}(F-2xF_x)=0, 
\label{evolutionphi}
\end{equation}
with $'=d/d\phi$.

For simplicity, we consider the $k$-field evolving linearly with time \cite{Feinstein:2002aj,Chimento:2009hg}; i.e. $\phi=\phi_0t$ where $\phi_0=\rm{constant}$.
Then, \mbox{$\alpha_0=F-2xF_x|_{x=-\phi_0^2}$}  is constant and Eq.~(\ref{evolutionphi}) implies
\begin{equation}
H=-\frac{\phi_0}{3\gamma_0}\frac{V'}{V},\qquad  \ga=\frac{2F_x\phi_0^2}{\alpha_0}=\gamma_0.
\label{hofV}
\end{equation}
The previous equation can be integrated with result $V=V_0a^{-3\gamma_0}$. We obtain the same potential deduced in \cite{Chimento:2009hg} because the result is independent of the modified Friedmann equation of the brane. Similarly, we can show that $\rho=\rho_0a^{-3\gamma_0}$, where $\rho_0=\alpha_0 V_0$.

In order to analyse the dynamics of the brane, it is useful to introduce the following variable 
\begin{equation}
Z=\frac{\rho_m}{\rho}.
\label{defZ}
\end{equation}
Therefore, the Hubble parameter can be expressed as
\begin{equation}
H=\frac{1}{3\gamma_0}\frac{\dot Z}{Z}.
\label{hofZ}
\end{equation}
Notice that by specifying the variable $Z$, we can determine the $k$-essence field potential as $V\propto 1/ Z$. This last relation can be deduced by combining Eqs.~(\ref{hofV}) and (\ref{hofZ}) and $\phi=\phi_0 t$. 
Finally, by substituting Eq.~(\ref{hofZ}) in the modified Friedmann equation (\ref{Friedmannroot}), we obtain 
\begin{equation}
\dot{Z}=\frac{3\epsilon\gamma_0 Z}{2r_c}\left[-1\pm\sqrt{1+\frac{1}{Z}}\right].
\label{dotZ}
\end{equation}

There are two aspects of the normal DGP branch that deserves to be commented: (i) it requires some sort of dark energy, for example in the form of a $k$-essence field, to describe the late-time behaviour of the universe, while   the self-accelerating branch does not require any cosmic fuel to mimic the current acceleration of the universe \cite{Durrer:2008in,Maartens:2010ar}; (ii) the normal branch is free of ghosts, a characteristic presents on the self-accelerating branch \cite{Koyama:2007za}.  

At this point, it is interesting to note that the two branches of the DGP model can be related by a form invariant transformation that preserve the form of the Friedmann and conservation equations. In fact, the change
\bn{t}
H\rightarrow -H, \quad \ep\rightarrow -\ep,  \quad \rho\rightarrow \rho,  \quad \rho+p\rightarrow -(\rho+p),
\ee
transforms the two branches of Eq. (\ref{Friedmann}) between them and preserves the form of the conservation equation (\ref{conservation}). This internal symmetry gives rise to a duality: $a\rightarrow 1/a$, which after integration implies $H \rightarrow -H$, between expanding and contracting universes \cite{Chimento:2003qy}. More precisely, if we know a solution $a$ of the branch $\ep$ corresponding to a fluid with energy density $\ro$ and pressure $p$, that satisfies the weak energy condition, then $1/a$ is a solution of the branch $-\ep$ with the same energy density and pressure $-2\ro-p$, that violates the weak energy condition. It means that there is a symmetry between the normal and the self-accelerating branches.

Along this paper, we will use the normal branch and the conclusion for the self-accelerating branch will be obtained by applying the form invariant transformation (\ref{t}). 

\section{The normal DGP branch}
\label{sec3}

In what follows, we concentrate our analysis on the solutions of a brane filled with the $k$-field introduced on the previous section for $\epsilon=1$, corresponding to the normal DGP branch geometry. There are four cases to be analysed depending on the sign of the equation of state $\gamma_0$ and the energy density of the $k$-field $\rho$. 

\subsection{Positive energy}

For $0\leq\rho$; i.e. $0\leq Z$, the Eq.~(\ref{dotZ}) can be integrated analytically \cite{W}
\begin{eqnarray}
&&Z\pm\left[\sqrt{{Z}^2+Z}+\frac12\ln\left(\frac12+Z+\sqrt{{Z}^2+Z}\right)\right] \nonumber\\
&&-K_1^\pm\,=\,\frac{3\gamma_0}{2r_c}(t-t_1), 
\label{Z1}
\end{eqnarray}
where $K_1^\pm$ and $t_1$ are constants. The solutions of Eq. (\ref{dotZ}) for $+$ and $-$ will be written as $Z^+$, $Z^-$ respectively.

If the $k$-field mimics a cosmological constant; i.e. $\gamma_0=0$, then Eq.~(\ref{Z1}) 
is easily satisfied as the lhs and rhs of Eq.~(\ref{Z1}) vanish (notice that the energy density and, therefore, the variable $Z^\pm$ are constants on this particular case). 

At high energy $\rho_m\ll \rho$, i.e. $Z^{\pm}\ll 1$ (see Eq.~(\ref{defZ})), $Z^\pm$ can be approximated as 
\begin{equation}
\sqrt{Z^\pm}\sim\pm\frac{3\gamma_0}{4r_c} t, 
\label{Zheprho}
\end{equation}
where we have made a rescaling of the cosmic time. Therefore, $V^\pm\propto (\phi_0/\phi)^2$. We kept the square root on the previous equation because it is crucial to fix the sign of the cosmic time of the brane for a given sign choice of $\gamma_0$. At low energy $\rho\ll\rho_m$; i.e. $1\ll Z^\pm$, it can be proved that
\begin{eqnarray}
Z^+&=&\frac{3\gamma_0}{4r_c} t, \label{Zleprho1}\\
\ln(Z^-)&=&-\frac{3\gamma_0}{r_c} t \label{Zleprho2}.
\end{eqnarray}
Consequently, the $k$-field potential on this regime fulfils: $V^+\propto\phi_0/\phi$ and $V^-\propto\exp[3\gamma_0\phi/(r_c\phi_0)]$. 
Let us point out also the high energy regime corresponds to a 4d regime; i.e. $H^2\sim\kappa_4^2\rho/3$.
To complete our analysis, based on Eqs.~(\ref{Z1}),~(\ref{Zheprho}),~(\ref{Zleprho1}) and (\ref{Zleprho2}),  is it is helpful to distinguish two cases: a positive and a negative $\ga_0$. 

\subsubsection{Positive $\ga_0$}
 
The solution $Z^+$ corresponds to a brane that starts its evolution with a Big Bang singularity where  $\rho\sim\ 4/(3\gamma_0^2\kappa_4^2) t^{-2}$. Consequently, at early time, coinciding with the high energy regime,  the scale factor scales as $a\propto t^{2/(3\gamma_0)}$ corresponding to the relativistic behaviour of a 4d Friedmann-Lema\^
{\i}tre-Robertson-Walker (FLRW) universe filled with the same matter content as the brane. This is in agreement with the fact that a homogeneous and isotropic  DGP brane at high energy behaves like a FLRW universe in 4d general relativity \cite{dgpfriedmann,BouhmadiLopez:2004ys}. The brane keeps expanding and at very late-time, the energy density and scale factor can be approximated by $\rho\sim 1/(\kappa_4^2r_c\gamma_0) t^{-1}$ and $a\propto t^{1/(3\gamma_0)}$, respectively. This is a consequence of the low energy behaviour of the normal DGP branch (with a + sign in Eq.~(\ref{Friedmannroot})) where $H\sim r_c\kappa_4^2\rho/3$ \cite{dgpfriedmann,BouhmadiLopez:2004ys}.

The solution $Z_-$ is asymptotically de Sitter on the past where $\ro\sim\exp{(3\ga_0t/r_c)}$ and $a\propto\exp(-t/r_c)$. Notice that $t$ is negative as can be easily seen from Eq.~(\ref{Zheprho}) bearing in mind that $\gamma_0$ is positive. Afterwards the brane starts contracting until it hits a big crunch in the future, where $\rho\sim 4/(3\gamma_0^2\kappa_4^2)t^{-2}$ and $a\sim t^{2/(3\gamma_0)}$. Because the brane is contracting in this case and $0<\gamma_0$, the high energy regime takes place at late-time while the low energy regime describes the early-time evolution of the brane.

\subsubsection{Negative $\ga_0$}

For a negative $\ga_0$, the energy density is a growing function of the scale factor. Therefore, the high-energy regime corresponds to large scale factors while the low-energy regime takes place at small  scale factors.

The brane described by $Z^+$ starts its expansion at $t\rightarrow -\infty$ where $a\sim 0$.  More precisely, at early time $\rho\sim 1/(\kappa_4^2r_c\gamma_0) t^{-1}$ and $a^{3\gamma_0}\propto \gamma_0t$. This regime corresponds to the low-energy regime of the normal DGP branch (with a + sign in Eq.~(\ref{Friedmannroot})) where the Hubble rate is approximately a linear function of the energy density of the brane. The brane keeps expanding in a super-accelerating way; i.e. $0<\dot H$, as Eq.~(\ref{Raychaudhuri}) implies for $\gamma_0<0$, $0<\rho$ and $\epsilon=1$. The expansion of the brane halts when it hits a big rip singularity at $t=0$ where the scale factor, the energy density, the pressure, the Hubble rate and its cosmic derivative blow up \cite{Caldwell:1999ew,Nojiri:2005sx}. The last stage of the brane expansion corresponds to the high energy regime; therefore a 4d regime, where $\rho\sim\ 4/(3\gamma_0^2\kappa_4^2) t^{-2}$ and $a^{3\gamma}\propto t^2$ while $t\rightarrow 0^-$.

On the other hand, the brane corresponding to $Z^-$ starts its evolution with a type I singularity \cite{Nojiri:2005sx} in the past. Initially, the energy density and scale factor are very large, indeed $\rho\sim\ 4/(3\gamma_0^2\kappa_4^2) t^{-2}$ and $a^{3\gamma}\propto t^2$ where $t\rightarrow 0^+$. This corresponds to the high energy regime. Then the brane starts contracting; i.e. $H<0$, even though $0<\ddot a$ because $0<\dot H$. The brane is asymptotically de Sitter in the future, corresponding to the low energy regime, where  $\rho\sim \rho_m\exp(3\gamma_0t/r_c)$ and $a\propto \exp(-t/r_c)$. 

\subsection{Negative energy density}

The DGP model has solutions with a $k$-field with a finite negative energy density as long as \mbox{$-\rho_m\leq\rho< 0$}\,; i.e.  $Z\leq-1$, cf. Eqs.~(\ref{Friedmannroot}) and (\ref{dotZ}). At the minimum energy density $\rho=-\rho_m$, which corresponds to a finite value of the scale factor, the Hubble rate vanishes (see Eq.~(\ref{Friedmannroot})), while the derivative of the Hubble rate diverges (cf. Eq.~(\ref{Raychaudhuri})), eventhough the energy density and the pressure are finite. Therefore this is a ``quiescent" singularity \cite{Shtanov:2002ek}, but notice that it is different from the one discussed in \cite{Shtanov:2002ek} which was induced by: 
\begin{itemize}
\item the presence of a dark radiation term on the brane, and therefore a black hole in the bulk which is not the case in the DGP model, and matter with positive energy density and whose equation of state satisfies specific conditions.
\item or through an inequality  condition satisfied by a linear combination of the bulk cosmological constant and the brane tension which cannot be fulfilled  on the DGP model.
\end{itemize}
Note as well  that a ``quiescent" singularity can also shows up in a DGP brane-world model with a Gauss-Bonnet term in the bulk \cite{Kofinas:2003rz}. Here again, the singularity is different from the one that takes place at $\rho=-\rho_m$. The singularity in \cite{Kofinas:2003rz} is caused by a combination of infra-red and ultra-violet modifications of general relativity. The presence of this singularity is the main motivation for looking for the solution of Eq.~(\ref{dotZ}) when the energy density is negative.  We summarise next the dynamics of the normal branch on those cases.

For $\rho \leq 0$; i.e. $Z^{\pm}\leq 0$, Eq.~(\ref{dotZ}) can be rewritten as 
\begin{equation}
\dot{Z}=\frac{3\gamma_0}{2r_c}\left[-Z\mp\sqrt{{Z}^2+Z}\right],
\label{dotZ3}
\end{equation}
which can be also integrated analytically \cite{W}
\begin{eqnarray}
&&Z\mp\left[\sqrt{{Z}^2+Z}+\frac12\ln\left(-\frac12-Z-\sqrt{{Z}^2+Z}\right)\right] \nonumber\\
&&-K_2^\pm\,=\,\frac{3\gamma_0}{2r_c}(t-t_2), 
\label{Z2}
\end{eqnarray}
where $t_2$ and $K_2^\pm$ are constants. Here again, the function $Z^+$, $Z^-$ will refer to the $Z$ function satisfying Eq.~(\ref{dotZ}) with $+$ and $-$, respectively.

For $\rho \leq 0$ with $\ro_0<0$; i.e. $Z^{\pm}\leq 0$, the solution (\ref{Z2}) can be expanded about the minimum energy density, $\rho=-\rho_m$ or $Z=-1$, 
\begin{equation}
Z^\pm+1\sim\frac{3\gamma_0}{2r_c}t \pm \frac23 \left(-\frac{3\gamma_0}{2r_c}t\right)^{\frac32},
\label{Zsnrho}
\end{equation}
where $a_m^{3\ga_0}=-\ro_0/\ro_m$. Below, we will see that the second term in the expansion is responsible for the divergence of  $\ddot a$ when it is evaluated at $a=a_m$ or $t=0$. At this point the brane approaches the ``quiescent" singularity because $H$ remains finite but $\dot H$ diverges. For simplicity, we have set $a=a_m$ at $t=0$. The potential close to the singularity can be approximated by $V^\pm\sim V_0 a_m^{-3\gamma_0} [1-3\gamma_0\phi/(2r_c\phi_0)]^{-1}$ and it is clearly finite at the singularity where $\phi=0$. The scale factor $a_m$ stands for the scale factor at the ``quiescent" singularity which is the minimum scale factor of the brane for $0<\gamma_0$ and the maximum scale factor of the brane for  $\gamma_0<0$.

On the other hand, at very low energy ($\rho\rightarrow 0^-$ or $Z^\pm \rightarrow -\infty$), we obtain 
\begin{eqnarray}
Z^+&\sim&\frac{3\gamma_0}{4r_c} t, \label{Zlenrho1}\\
\ln(-Z^-)&\sim&-\frac{3\gamma_0}{r_c} t. \label{Zlenrho2}
\end{eqnarray}
The $k$-field potential on this regime fulfils: $V^+\propto\phi_0/\phi$ and $V^-\propto\exp[3\gamma_0\phi/(r_c\phi_0)]$. 
It is also worthy to mention that the brane always shrinks when filled by a negative energy density (see Eq.~(\ref{Friedmannroot}) with $\epsilon=1$).
Like in the previous subsection, to complete our analysis, based on Eqs.~(\ref{Z1}),~(\ref{Zsnrho}),~(\ref{Zlenrho1}) and (\ref{Zlenrho2}),  is it is helpful to distinguish two cases: a positive and a negative $\ga_0$. 

\subsubsection{Positive $\ga_0$}

For $Z^+$ and $0<\gamma_0$, the brane starts its evolution with a very large radius and an almost vanishing energy density, where $\rho\sim \ro_0 r_c/t$ and $a\propto (t/r_c)^{1/3\gamma_0}$. The Hubble rate and the derivative of the Hubble rate are almost zero initially. This solution is defined in a region where the cosmic time is negative. Afterwards the brane starts contracting until it hits the ``quiescent" singularity at a minimum scale factor 
\ben
\n{1}
&&\rho\sim -\rho_m\left[1+\frac{3\gamma_0t}{2r_c}+\frac23\left(-\frac{3\gamma_0t}{2r_c}\right)^{3/2}\right],\\
&&a\sim a_m\left[1-\frac{t}{2r_c}-\frac{2}{9\gamma_0}\left(-\frac{3\gamma_0t}{2r_c}\right)^{3/2}\right],
\een
where the latter equation shows explicitly that $\ddot a$ diverges at $t=0$.

The solution corresponding to $Z^-$ is asymptotically de Sitter in the past where once more the scale factor is very large and the energy density is approaching zero. In more specific terms, we obtain $\rho\sim -\rho_m\exp(3\gamma_0t/r_c)$ and $a\propto\exp(-t/r_c)$ where $t\ll -1$. Even though the  brane is asymptotically de Sitter in the past, it starts shrinking because the Hubble rate is initially (almost) constant and negative. In the future the brane face a ``quiescent" singularity at a minimum scale factor characterised by 
\ben
&&\rho\sim -\rho_m\left[1+\frac{3\gamma_0t}{2r_c}-\frac23\left(-\frac{3\gamma_0t}{2r_c}\right)^{3/2}\right],\\
\n{2}
&&a\sim a_m\left[1-\frac{t}{2r_c}+\frac{2}{9\gamma_0}\left(-\frac{3\gamma_0t}{2r_c}\right)^{3/2}\right],
\een
and again $\ddot a$ diverges at $t=0$.

\subsubsection{Negative $\ga_0$}

The solution $Z^{\pm}$ starts at a ``quiescent" singularity, where the brane reaches its maximum scale factor $a_m$, and has the same behaviour described by the solutions (\ref{1})-(\ref{2}). In this case the solutions are defined in the region where the brane cosmic time is positive. The solutions $Z^\pm$ describe a contracting brane that ends with a vanishing scale factor even though at different rates. For $Z^+$, the energy density and scale factor change with the cosmic time as $\rho\sim \ro_0 r_c/t$ and $a\propto (t/r_c)^{1/3\gamma_0}$,  while for $Z^-$, the same quantities behaves as $\rho\sim -\rho_m\exp((3\gamma_0t)/r_c)$ and $a\propto\exp(-t/r_c)$. In both cases, the brane shrinks to a point in an infinite amount of its proper time.

\section{Conclusions}

We have investigated a DGP brane filled with a $k$-essence field. This source is particularly interesting because it can take negative values and gives the possibility of considering ``quiescent'' singularities in the DGP framework. In addition, it gives rise to inflationary branes. We have focused our analysis on the normal DGP branch, for the reasons explained in the introduction, fundamentally based on the duality between both branches as we have shown at the end of section II . Our results for the normal branch are summarised on the next paragraphs.

In the $0<\rho$ case we have found: (i) a singular scale factor that represents an expanding universe with a final power law scenario and, a  contracting one with an initial de Sitter phase and a final big crunch at $t=0$ for $\gamma_0>0$. However, for $\gamma_0<0$, (ii) we have found an expanding scale factor, such that, the universe begins with a vanishing value at $t=-\infty$ and ends in a big rip singularity at $t=0$. There exists also a singular universe which contracts from $a=\infty$, at $t=0$, to an asymptotically de Sitter scenario in the future.

In the $\ro<0$ case we have obtained a contracting scale factor for any value of the equation of state $\gamma_0$. For $\gamma_0>0$ the universe contracts from the far past at $t=-\infty$ with a power law expansion or a de Sitter phase and ends with a constant scale factor $a_m$, at $t=0$. We have shown that  the expansion rate is finite; more precisely $H=H_m=-1/2r_c$, at $t=0$ and $\dot H_m$ diverges, indicating that the universe ends in a quiescent singularity \cite{Shtanov:2002ek}.  Nevertheless the potential becomes finite when it is evaluated at this singularity, $V_m=V_0 a_m^{-3\gamma_0}$. Similarly, the energy density and the pressure of the $k$-field are finite at this point. For $\gamma_0<0$ the universe begins from the constant value $a_m$ in a ``quiescent'' singularity, after that, it contracts and ends with a power law or a de Sitter phases.

\section*{Acknowledgments}

MBL is  supported by the Portuguese Agency Funda\c{c}\~{a}o para a Ci\^{e}ncia e
Tecnologia through the fellowship SFRH/BPD/26542/2006. She also wishes
to acknowledge the hospitality of LeCosPA at the National University of Taiwan and the University of the Basque Country
during the completion of part of this work. LPC thanks the hospitality of the University of the Basque Country, the Basque Foundation for Science (Ikerbasque), the University of Buenos Aires under Project No. X044 and the Consejo Nacional de Investigaciones Cient\'{\i}ficas y T\' ecnicas (CONICET)under Project PIP 114-200801-00328 for the partial support of this work during their different stages.

\end{document}